\documentclass[manuscript]{aastex}

\shorttitle{SEPs in the Earth's Magnetosheath}
\shortauthors{Adriani et al.}

\begin{document}


\title{PAMELA's Measurements of Magnetospheric Effects on High Energy Solar Particles}




\author{
O.~Adriani$^{1,2}$,
G.~C.~Barbarino$^{3,4}$,
G.~A.~Bazilevskaya$^{5}$,
R.~Bellotti$^{6,7}$,
M.~Boezio$^{8}$,
E.~A.~Bogomolov$^{9}$,
M.~Bongi$^{1,2}$,
V.~Bonvicini$^{8}$,
S.~Bottai$^{2}$,
U.~Bravar$^{10}$,
A.~Bruno$^{6}$,
F.~Cafagna$^{7}$,
D.~Campana$^{4}$,
R.~Carbone$^{8}$,
P.~Carlson$^{14}$,
M.~Casolino$^{11,15}$,
G.~Castellini$^{16}$,
E.~C.~Christian$^{17}$,
C.~De Donato$^{11,12}$,
G.~A.~de Nolfo$^{17,**}$,
C.~De Santis$^{11}$,
N.~De Simone$^{11}$,
V.~Di Felice$^{11,20}$,
V.~Formato$^{8,18}$,
A.~M.~Galper$^{13}$,
A.~V.~Karelin$^{13}$,
S.~V.~Koldashov$^{13}$,
S.~Koldobskiy$^{13}$,
S.~Y.~Krutkov$^{9}$,
A.~N.~Kvashnin$^{5}$,
M.~Lee$^{10}$,
A.~Leonov$^{13}$,
V.~Malakhov$^{13}$,
L.~Marcelli$^{12}$,
M.~Martucci$^{12,21}$,
A.~G.~Mayorov$^{13}$,
W.~Menn$^{19}$,
M.~Merg$\acute{e}$$^{11,12}$,
V.~V.~Mikhailov$^{13}$,
E.~Mocchiutti$^{8}$,
A.~Monaco$^{6,7}$,
N.~Mori$^{1,2}$,
R.~Munini$^{8,18}$,
G.~Osteria$^{4}$,
F.~Palma$^{11,12}$,
B.~Panico$^{4}$,
P.~Papini$^{2}$,
M.~Pearce$^{14}$,
P.~Picozza$^{11,12}$,
M.~Ricci$^{21}$,
S.~B.~Ricciarini$^{2,16}$,
J.~M.~Ryan$^{10}$,
R.~Sarkar$^{22,*}$,
V.~Scotti$^{3,4}$,
M.~Simon$^{19}$,
R.~Sparvoli$^{11,12}$,
P.~Spillantini$^{1,2}$,
S.~Stochaj$^{23}$,
Y.~I.~Stozhkov$^{5}$,
N.~Thakur$^{17}$,
A.~Vacchi$^{8}$,
E.~Vannuccini$^{2}$,
G.~I.~Vasilyev$^{9}$,
S.~A.~Voronov$^{13}$,
Y.~T.~Yurkin$^{13}$,
G.~Zampa$^{8}$,
and N.~Zampa$^{8}$
}

\affil{$^{1}$ Department of Physics and Astronomy, University of 
Florence, I-50019 Sesto Fiorentino, Florence, Italy}
\affil{$^{2}$ INFN, Sezione di Florence, I-50019 Sesto Fiorentino, 
Florence, Italy}
\affil{$^{3}$ Department of Physics, University of Naples ``Federico 
II'', I-80126 Naples, Italy}
\affil{$^{4}$ INFN, Sezione di Naples, I-80126 Naples, Italy}
\affil{$^{5}$ Lebedev Physical Institute, RU-119991 Moscow, Russia}
\affil{$^{6}$ University of Bari, I-70126 Bari, 
Italy}
\affil{$^{7}$ INFN, Sezione di Bari, I-70126 Bari, Italy}
\affil{$^{8}$ INFN, Sezione di Trieste, I-34149 Trieste, Italy}
\affil{$^{9}$ Ioffe Physical Technical Institute, RU-194021 St. 
Petersburg, Russia}
\affil{$^{10}$ Space Science Center, University of New Hampshire, Durham, NH, USA.}
\affil{$^{11}$ INFN, Sezione di Rome ``Tor Vergata'', I-00133 Rome, Italy}
\affil{$^{12}$ Department of Physics, University of Rome ``Tor 
Vergata'', I-00133 Rome, Italy}
\affil{$^{13}$ National Research Nuclear University MEPhI, RU-115409 
Moscow, Russia}
\affil{$^{14}$ KTH, Department of Physics, and the Oskar Klein Centre for Cosmoparticle Physics, AlbaNova University Centre, SE-10691 Stockholm, Sweden.}
\affil{$^{15}$ RIKEN, Advanced Science Institute, Wako-shi, Saitama, Japan}
\affil{$^{16}$ IFAC, I-50019 Sesto Fiorentino, Florence, Italy}
\affil{$^{17}$ Heliophysics Division, NASA Goddard Space Flight Ctr, Greenbelt, MD, USA.}
\affil{$^{18}$ University of Trieste, Department of Physics, I-34147 Trieste, Italy.} 
\affil{$^{19}$ Universit at Siegen, Department of Physics, D-57068 Siegen, Germany.}
\affil{$^{20}$ Agenzia Spaziale Italiana (ASI) Science Data Center, Via del Politecnico snc I-00133 Rome, Italy.}
\affil{$^{21}$ INFN, Laboratori Nazionali di Frascati, Via Enrico Fermi 40, I-00044 Frascati, Italy. }
\affil{$^{22}$ Indian Centre for Space Physics, 43, Chalantika, Garia Station Road, Kolkata 700 084, West Bengal, India.}
\affil{$^{23}$ Electrical and Computer Engineering, New Mexico State University, Las Cruces, NM, USA.}
\affil{$^{*}$ Previously at INFN, Sezione di Trieste, I-34149 Trieste, Italy.}
\affil{$^{**}$ Correspondence to:  G. A. de Nolfo, georgia.a.denolfo@nasa.gov. }

\begin{abstract}
The nature of particle acceleration at the Sun, whether through flare reconnection processes or through shocks driven by coronal mass ejections (CMEs), is still under scrutiny despite decades of research.  The measured properties of solar energetic particles (SEPs) have long been modeled in different particle-acceleration scenarios.  The challenge has been to disentangle to the effects of transport from those of acceleration. The Payload for Antimatter Matter Exploration and Light-nuclei Astrophysics (PAMELA) instrument, enables unique observations of SEPs including composition and the angular distribution of the particles about the magnetic field, i.e. pitch angle distribution, over a broad energy range ($>$80 MeV) -- bridging a critical gap between space-based measurements and ground-based.  We present high-energy SEP data from PAMELA acquired during the 2012 May 17 SEP event.  These data exhibit differential anisotropies and thus transport features over the instrument rigidity range.  SEP protons exhibit two distinct pitch angle distributions; a low-energy population that extends to 90$^{\circ}$ and a population that is beamed at high energies ($>$ 1 GeV), consistent with neutron monitor measurements. To explain a low-energy SEP population that exhibits significant scattering or redistribution accompanied by a high-energy population that reaches the Earth relatively unaffected by dispersive transport effects, we postulate that the scattering or redistribution takes place locally. We believe these are the first comprehensive measurements of the effects of solar energetic particle transport in the Earth's magnetosheath.
\end{abstract}





\section{Introduction}
Whether the Sun accelerates particles at low altitudes through magnetic reconnection or higher in the corona through Coronal Mass Ejection-driven shocks, or perhaps an admixture of the two, is unclear.  The uncertainty exists even at the very lowest energies of solar energetic particles (SEPs) observed in-situ and persists right through Ground Level Enhancements (GLEs). The challenge confronting researchers is that the signatures of acceleration are modified or obscured as a consequence of transport within interplanetary space.  The complexion of SEP events measured in space often departs markedly from that registered on the ground, begging the question of whether we are witnessing two separate processes or strong and energy sensitive transport effects (e.g. \citet{Mathews1990, Vashenyuk2006, McCracken2008}). 

Besides confounding the study of particle acceleration, the physics of energetic particle transport within the heliosphere is its own topic of study.  The highest energy SEPs (observed in GLEs) provide an idealized case study in that the effects of transport are often minimal.  How the different mechanisms of transport, however, translate to lower energies is poorly understood.  The unique observations from the Payload for Antimatter Matter Exploration and Light-nuclei Astrophysics (PAMELA) instrument, a low-Earth satellite with sensitivity to composition over a broad range in energy (80 MeV to several GeV), provide an essential link between the highest energy GLEs and the low-energy direct observations of spacecraft within the inner heliosphere.  

\section{The 2012 May 17 GLE}

PAMELA observed the first Ground Level Enhancement (GLE) of solar cycle 24, the GLE of 2012 May 17, offering a rare opportunity to directly measure solar energetic particles from 80 MeV to several GeV.  Significant count rates ($\ge$15\% over background) were first registered by the Apatity neutron monitor at 0150 UT (Papaioannou et al. 2014).  Neutron monitors (NMs) at Oulu and Mawson also registered the event with signals $\ge$15\% over background.  All three NMs had viewing angles (or computed asymptotic directions) close to the interplanetary magnetic field (IMF) direction, in agreement with the computations of \citet{Mishev2013} and \citet{Papaioannou2014}.  South Pole and Polar Bare NMs as well as ICETop \citep{Ahrens2013} also registered a prompt signal despite having viewing directions farther from the direction of the IMF.  The initial prompt signal was followed by a broader enhancement that was registered in all of the above NMs but also in other stations at the few-percent level.  

The GLE was associated with an M5.1 flare in active region AR11476 located on the Sun at N07W88.  The LASCO/SOHO measured an associated fast CME with a maximum speed of 1997 km/s \citep{Gopalswamy2013}. Type II radio emission was reported by the Solar Geophysical Data at 0131 UT signifying the formation of a shock.  The computed injection time at the Sun using NM data, WIND helium data, and an upper limit from PAMELA is 0139 $\pm$ 1min UT \citep{Thakur2012}, consistent with the onset time computed by \citet{Gopalswamy2013}.  The injection was delayed by ~8 minutes from the formation of the shock as delineated by Type II radio emission  \citep{Gopalswamy2013}.  The presence of accelerated high-energy ions at the Sun (whether these ions escape to be registered as GLEs is not clear) is evidenced by  $>$ 100 MeV gamma-ray emission \citep{Ajello2014}.

However, even at GLE energies ( $>$ 1 GeV) an evolving pitch angle distribution results from transport effects that are complex and complicate the interpretation of the data.  GLEs often consist of a rapid, anisotropic onset associated with a well-connected beam of anti-sunward, outward moving particles aligned with the IMF.  This initial onset has been interpreted as a result of magnetic focusing with little scattering and predictable velocity dispersion \citep{Earl1976}.  As the event progresses, however, the distribution becomes isotropic and decays, presumably due to pitch-angle scattering, as the particles diffuse into the inner heliosphere. \citet{Fluckiger1993} and \citet{Cramp1997} proposed that this isotropic component is a result of scattering off a magnetic structure beyond Earth or perhaps is due to a magnetic bottleneck as proposed by \citep{Bieber2002}. Shea \& Smart considered two separate injections of particles at the Sun that propagate along the IMF \citep{SheaSmart1997,  Makhmutov2002}.  However, bi-directional flows have been detected in neutron monitor data that suggest closed magnetic loops with footpoints at the Sun, commonly observed with solar energetic electrons.  The calculations by \citet{Ruffolo2006} of the 1989 October 22 GLE support particle transport (and scattering) within a large (1 AU) closed magnetic loop structure. 

The shape and morphology of GLEs is usually determined through extensive modeling of the response of the world-wide network of neutron monitors located at widely differing locations and altitudes.  The directional distribution of the incoming event is modeled to determine the omnidirectional density and weighted anisotropy (\citet{Bieber2013}, and references therein).  While there has been success in modeling the gross features of the incoming GLE beam, this method depends on the neutron monitor yield function and the assumption of vertical asymptotic directions.  Neutron monitors have no energy discrimination, but an energy dependent field of view. Their count rate reflects the particle intensity with energies above geomagnetic cutoff integrated over a large solid angle.  

To fully understand the extent of the anisotropy and its evolution requires measurements over a broad range in energy, and also the ability to determine directly the pitch angle at energies above 1 GeV, corresponding to the energy range of high latitude neutron monitors.  

\section{PAMELA Observations}

PAMELA is a space-based cosmic-ray detector built by the European/Russian WiZard collaboration launched into orbit aboard a Soyuz TM2 rocket on 2006 June 15. PAMELA is attached to the upward looking mounting point on the Russian Resurs-DK1 Earth-observing satellite.  In 2010, the spacecraft was maneuvered into a high inclination (70$^{\circ}$) circular orbit at an altitude of about ~580 km.  The instrument is comprised of a magnet spectrometer with a silicon tracking system and a time-of-flight system shielded by an anticoincidence system and underneath  an imaging Si-calorimeter and neutron detector \citep{Picozza2007, Adriani2013b}.   

PAMELA has made significant contributions to our knowledge of galactic cosmic-ray protons, electrons, and their antiparticles and light nuclei in the energy range 10$^2$-10$^6$ MeV \citep{Picozza2007, Adriani2009a, Adriani2009b, Adriani2011a, Adriani2011b, Adriani2013a}.  Details of the instrument performance, particle selection, and selection efficiencies were described by \citep{Adriani2013b}.  Its ability to measure the spectra, composition, and differential intensity over a broad range in energy offers opportunities for investigations of SEPs. PAMELA measures the incident trajectory of the detected particles employing a combination of trajectory reconstruction with the silicon tracking system and particle tracing techniques.  Its field of view is narrow $\sim$ 20$^{\circ}$.  Because it is a moving platform, it sweeps through pitch angle space allowing one to construct a pitch angle distribution of the SEPs.  For this study, we employed the particle tracing techniques of \citet{SheaSmart2000} to determine the asymptotic direction for each incident particle with respect to the IMF. Our tracing techniques \citep{Bruno2014} have been extensively tested with trapped and albedo protons \citep{Bruno2013} and are based on a realistic description of the Earth's magnetosphere, relying on an internal field model IGRF-2011 \citep{Finlay2010} and an external field model TS07 \citep{Tsyganenko2007, Sitnov2008}.  Solar wind and IMF parameters were obtained from the high-resolution Omniweb database (omniweb.gsfc.nasa.gov).  

PAMELA scans a narrow swath of pitch angle along its orbit.  While neutron monitors measure the particle intensity for a given pitch angle (above the local cutoff rigidity), PAMELA is a moving observatory, with a sensitivity to both energy and a pitch angle that depends on spacecraft location.  The ability to measure particle intensity with varying sensitivity to pitch angle over a large pitch angle and energy range is new.     

\section{Results}

Figure 1 shows the computed asymptotic vertical (look) direction for PAMELA during the polar pass that first registered the 2012 May 17 event.  Color coding reflects the particle rigidity ranging from 0.4 GV to 2.4 GV.  The gray solid curve shows the spacecraft trajectory for periods when PAMELA is at geomagnetic locations where the St\"ormer vertical cutoff is lower than 390 MV \citep{Stormer1950, Shea1965}. The direction of the interplanetary magnetic field obtained from the Omniweb database is shown as the black cross. Contours of constant pitch angle relative to the direction of the IMF are also shown.   As noted above, PAMELA is sensitive to a small range in pitch angle distribution that varies along the orbit and samples the pitch angle distribution from 0$^{\circ}$ to about 140$^{\circ}$ during the twenty-minute polar pass.  

Figure 2 (top panel) shows the pitch-angle distribution for which PAMELA is sensitive as a function of time in two rigidity ranges.  The bottom panel shows the particle intensity (corrected for background) in two rigidity ranges observed by PAMELA.  The intensity was computed as a function of pitch angle and rigidity, by correcting registered proton counts for the instrument efficiencies, livetime, and gathering power of the apparatus at each orbital position, taking into account the anisotropic flux exposition and the change in look directions along the orbits \citep{Bruno2014}.   The background intensity (black dashed curve), due to galactic cosmic rays (GCRs), was estimated using data registered two days prior to the event.  Within a given pitch angle distribution, the GCR background is isotropic, as expected.  Comparing this top panel with the bottom panel, we can see how the proton fluxes varied as a function of pitch angle (and thus time).  During times when PAMELA is sensitive to pitch angles $<$ 90$^{\circ}$ degrees, we observe an increase in the flux of 0.55-0.76 GV particles. Furthermore, when PAMELA is beyond 90$^{\circ}$ (shaded region), we see the 0.55-0.76 GV particle return to background, reflecting the passage of PAMELA into and out of a low-energy component confined to the forward pitch-angle hemisphere.  At high rigidities ($>$ 1 GV) the flux is constant for pitch angles greater than ~40$^{\circ}$, and increases markedly at small angles (note the logarithmic scale).  PAMELA observes the event $\sim$10 minutes after the Apatity onset (0150 UT) and continued to observe the event during the rising phase of the prompt emission as viewed by Apatity, Oulu, Mawson, and South Pole.

The dependence on pitch angle for three different energy regimes is shown in Figure 3.  PAMELA observes two populations simultaneously with very different pitch angle distributions.  We see a low-energy component (0.39-1.07 GV) confined to pitch angles $<$ 90$^{\circ}$ and a high-energy component (1.50-2.09 GV) that is beamed with pitch angles $<$ 30$^{\circ}$, consistent with neutron monitor observations (Figure 3, bottom panel).  The component with intermediate energies (1.07-1.50 GV) suggests a transition between the low and high energies, exhibiting a peak at small pitch angles and a cutoff at 90$^{\circ}$.  At rigidities $>$ 1 GV, corresponding to NM data, the particles are mostly field aligned.  Assuming symmetry about the IMF, this would suggest a beam full width of $\sim$40-60$^{\circ}$. Given the nature of the PAMELA observations, it is difficult to understand if or how these populations evolve during the $\sim$20-minutes polar pass, however it is striking that two distinct populations are present simultaneously relatively soon after the onset of the event as registered by the neutron monitors.  

\section{Discussion}

GLEs typically become isotropic with time, however, this is the first time that a GLE, shortly after onset, has been observed where the anisotropic component is accompanied by a broader component that has presumably undergone significant scattering or redistribution.  We believe that the scattering responsible for the broader distribution at lower rigidities must take place locally as implied by the time coincidence of the highly scattered particles and the beamed particles.  To understand this, we note that the transit time for a zero-pitch-angle proton (which will travel over 1.2 AU) of rigidity 400 MV (the lowest energy observed by PAMELA in this work) is $\sim$1400 s.  For a pitch angle of 60$^{\circ}$ the same proton would take 2800 s.  This should be compared to the transit time for a 1 GV proton with a pitch angle of 20$^{\circ}$--890 s, a discrepancy of 1910 s.  This is a lower limit to the discrepancy since the prevailing solar wind speed was slower than usual, $\sim$365 km s$^{-1}$ (translating to larger path lengths than the nominal 1.2 AU).  For both populations to be present at Earth at the same time, they must have had similar transit times followed by local scattering to explain the presence of large pitch angle particles at low rigidity.  On the other hand, we can compare the expected arrival time discrepancy between 400 MV and 1 GV protons with the same 20$^{\circ}$ pitch angle.  The transit time for our original 400-MV proton at a pitch angle of 20$^{\circ}$ is $\sim$1490 s--a discrepancy of less than 600 s.  This difference for similar pitch angles is reasonable given the 7-minute offset from the GLE onset. However, the low-rigidity particles at much greater pitch angles are not compatible with these numbers.  Thus, these two pitch-angle distributions, narrow and broad, could not have co-existed at injection.

The transition from a forward hemisphere to field-aligned pitch angle distribution as one passes the 1 GV rigidity value is curious.  This value is just below the high latitude neutron monitor threshold and if it were a common phenomenon, it would help explain why the morphology of a GLE is so different from lower energy particle events measured in space \citep{Mewaldt2012} and references therein. The anisotropic nature of the $>$1-GV particles was present for at least 30 minutes, starting from the onset (0150 UT) as registered with the NM network, through the 1-GV exposure times with PAMELA at 0220 UT.  The scattered component was registered as early as 0157 UT on the leading edge of the well-connected NM signal representing an extended overlap in time of the two different populations, i.e., the hemispherical low-rigidity particles and the field aligned high-rigidity particles.  There are few candidate volumes or processes that are capable of producing the effect that are local.  The relatively sharp gradient in pitch angle space constrains the characteristics of the volume of space responsible for the differential scattering.

If we assume that the field-aligned distribution initially applied to a wide range of rigidities, covering the PAMELA range, then we must search for local agents that disperse or broaden that distribution preferentially at lower rigidities, while leaving the higher rigidity particles relatively unaffected.  Three choices emerge as possibilities, (1) upstream turbulence in the solar wind in the form of Alfv\'en waves, (2) the intense and chaotic fields in the Earth's bow shock itself and (3) the Earth's magnetosheath.  Looking at the characteristic gyroradius of particles upstream of the Earth's bow shock, we note that the gyroradius of a 700 MV proton in a 5-nT IMF is 74 R$_E$ or $1/3$ AU, not consistent with the dimensions of the local scattering volume that describes our results.  We also note that the upstream rms magnetic field, a measure of the turbulence present, did not exceed 1 nT for several days after the event while the IMF ranged between 5 and 10 nT (ACE Science Center; Farrugia, priv. comm.).  \Citet{Mishev2013} suggested that for this event there may be significant local scattering from an interplanetary structure of a previous CME that the Earth enters on May 16, but the small value of B$_{rms}$ provides little evidence to support this.

The intense turbulence in the Earth's bow shock is probably insufficient to produce the effect given the thinness of the shock, of order 20 km \citep{Schwartz2011}.  However, the magnetosheath possesses an intense magnetic field ($\ge$20 nT) with typically large-amplitude mirror mode instability fluctuations ($\ge$50\%) and is much thicker (several RE) \citep{Southwood1993} and references therein, (Farrugia, priv. comm.).  The diagram in Fig. 4 schematically shows the orientation of the IMF, the bow shock, magnetosphere, the particle asymptotic directions and the spacecraft trajectory of the PAMELA measurements.  The IMF is mostly tangential to the solar wind ($By \gg Bx$ or $Bz$) with the high-rigidity particles entering the magnetosphere on the dawn-side flank.  The lowest rigidity particles are detected near local midnight.  On the dawn side flank the magnetosheath is much thicker, of order 5 R$_E$ or greater.  In this sheath the fluctuations are in the form of mirror mode, non-propagating waves.  These convect down the flanks of the magnetosphere as quasi-static large amplitude structures with a spatial scale in the plasma rest frame of 1000-2000 km \citep{Gutynska2008}.  Depending on the point of entry into the magnetosheath and the phase of the gyration about the mean field, equal rigidity, field-aligned particles would take different paths, effectively dispersing in pitch angle when they enter the quieter magnetosphere and the instrument orbit.  Pitch angle diffusion, in the sense of scattering off Alfv\'en waves, would not be significant and the particles would continue a forward path and not be scattered backward unless there was mirroring taking place--not possible at these rigidities.  Smaller gyroradius (lower rigidity) particles would be more greatly affected.  Because the wave amplitude is comparable to the mean field, comparing the mean gyroradius to the magnetosheath thickness is a useful indicator of the magnitude of the effect one would expect.  For example, a 400 MV protons in the ambient 20-nT field has a gyroradius of $\sim$12 R$_E$, while a 1.5 GV proton has a gyroradius of $\sim$40 R$_E$.  These are encouraging numbers in that this would imply significant, but not too intense, dispersion in the 5-R$_E$ thick magnetosheath for the 1.5-GV GLE protons, but much greater dispersion for the lower rigidity particles. The distorted orientation of the IMF may be a key factor in the anisotropy effect observed in the particle intensities, because entry into the magnetosphere on the flank significantly increases the diffusive volume compared to the nominal 45$^{\circ}$ of the Archimedes spiral and because the quasi-perpendicular nature of the IMF is exactly what is needed to produce large mirror mode fluctuations.   

This is the first observation of differential dispersion as particles transit the magnetosheath.  Low energy solar energetic protons were registered within the Earth's magnetosheath \citep{Debrunner1984, Vlasova2011}, that qualitatively show similar anisotropies to the high-energy component.

\section{Summary }

PAMELA near-Earth observations of the 2012 May 17 GLE have identified a clear rigidity dependence of the energetic particle pitch angle distribution, suggesting that the low-energy component undergoes significant scattering or redistribution while the highest-energy SEPs reach the Earth unhindered by dispersive transport effects.  For both populations to arrive at Earth simultaneously and early in the same solar event, the scattering must take place locally.  We believe that this is the first observation of the effects of local transport on the anisotropic phase of the GLE as SEPs propagate through the Earth's magnetosphere. The ability to measure pitch angle distribution over a broad range in energy has dramatically changed our view of the transport of SEPs for this event and may with future observations change our views of SEP transport in general.  This type of analysis is only possible with the unique capability of the PAMELA instrument.

\acknowledgments

The authors would like to acknowledge support from NASA Heliophysics and Solar Research grant, the National Science Foundation (SHINE), the Italian Space Agency (ASI), Deutsches Zentrum f\"ur Luft-und Raumfahrt (DLR), the Swedish National Space Board, the Swedish Research Council, the Russian Space Agency (Roscosmos), and the Russian Science Foundation. We gratefully thank N. Tsyganenko and Dr. Charlie Farrugia for helpful discussions, and I. M. Sitnov and G. K. Stephens  for support in the use of the TS07D model.

\clearpage



\begin{figure}
\epsscale{0.8}
\plotone{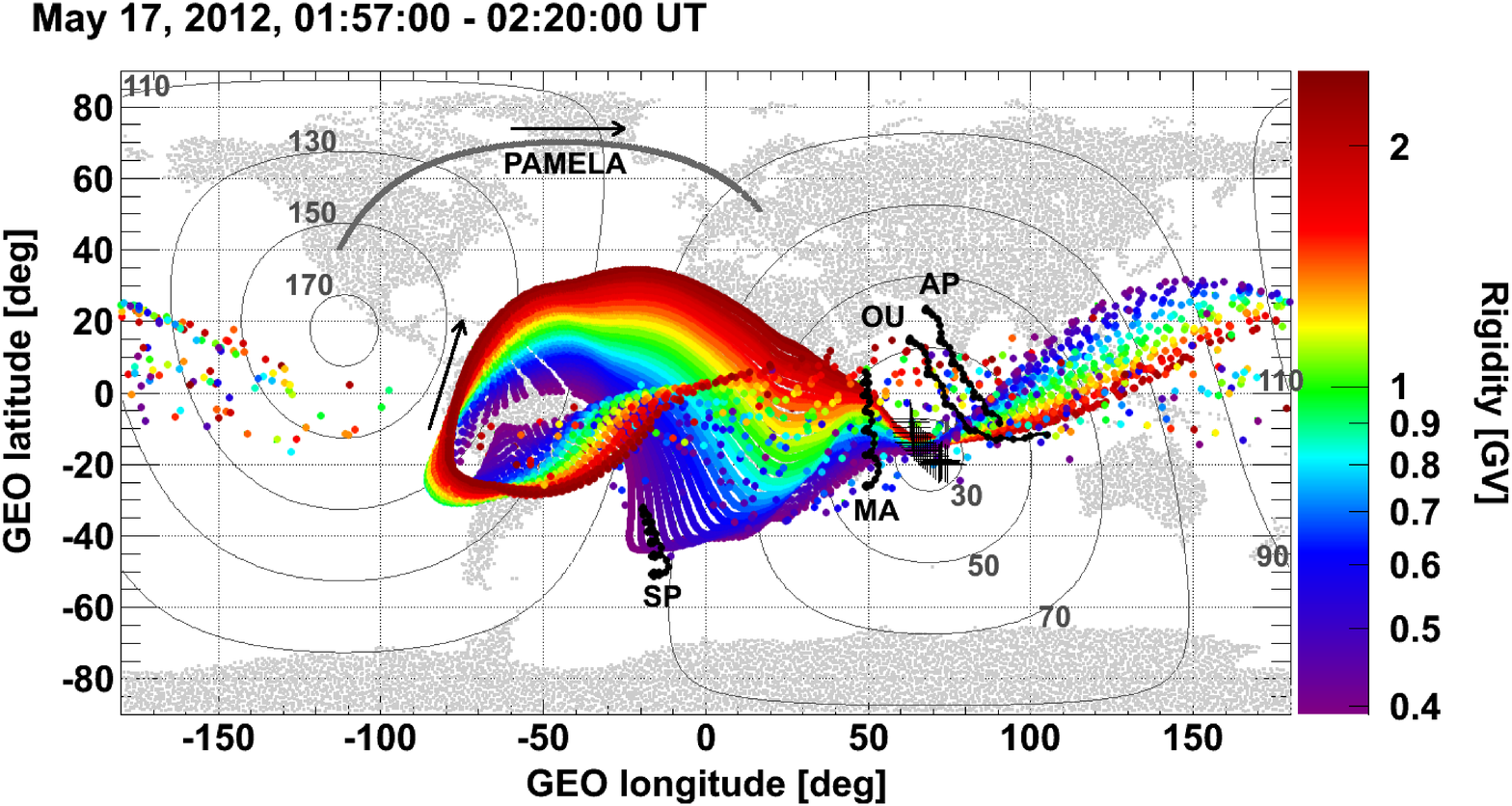}
\caption{Asymptotic directions determined during the first polar pass that registered the 2012 May 17 event. Also shown are the asymptotic directions of the NM that registered the primary GLE beam. OU, AP, SP, and MA stand for Oulu, Apatity, South Pole, and Mawson NMs, respectively. \label{fig1}}
\end{figure}

\clearpage

\begin{figure}
\epsscale{0.8}
\plotone{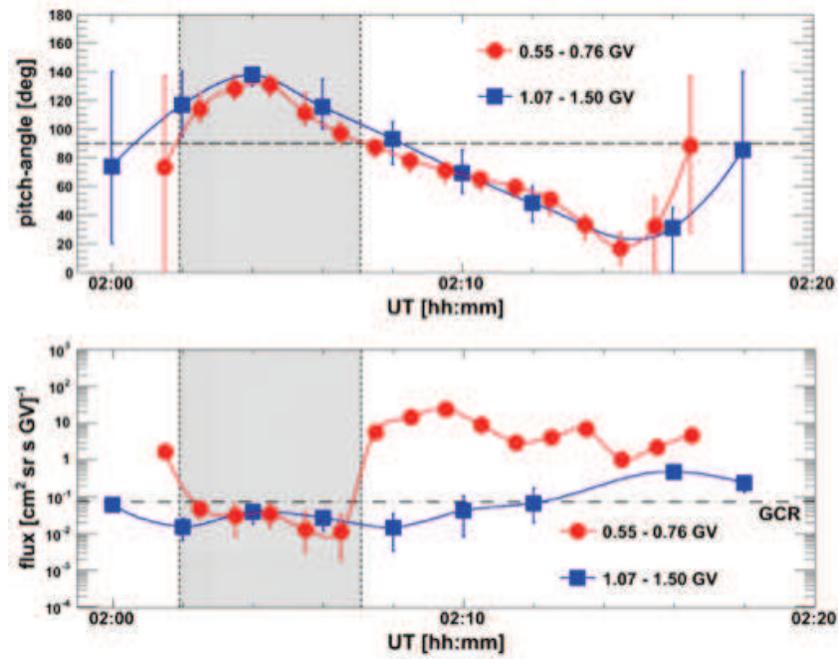}
\caption{(top) Mean pitch angle as a function of time during the polar pass for two rigidity ranges. The dashed line represents the 90$^{\circ}$ pitch angle. (bottom) Flux of SEPs registered by PAMELA in two rigidity ranges.  The shaded area represents the time during which PAMELA is above 90$^{\circ}$ pitch angle. GCR background (at 1.5 GV) is indicated as a dashed line. The curves through the data are meant to guide the eye.   \label{fig2}}
\end{figure}

\clearpage

\begin{figure}
\epsscale{0.8}
\plotone{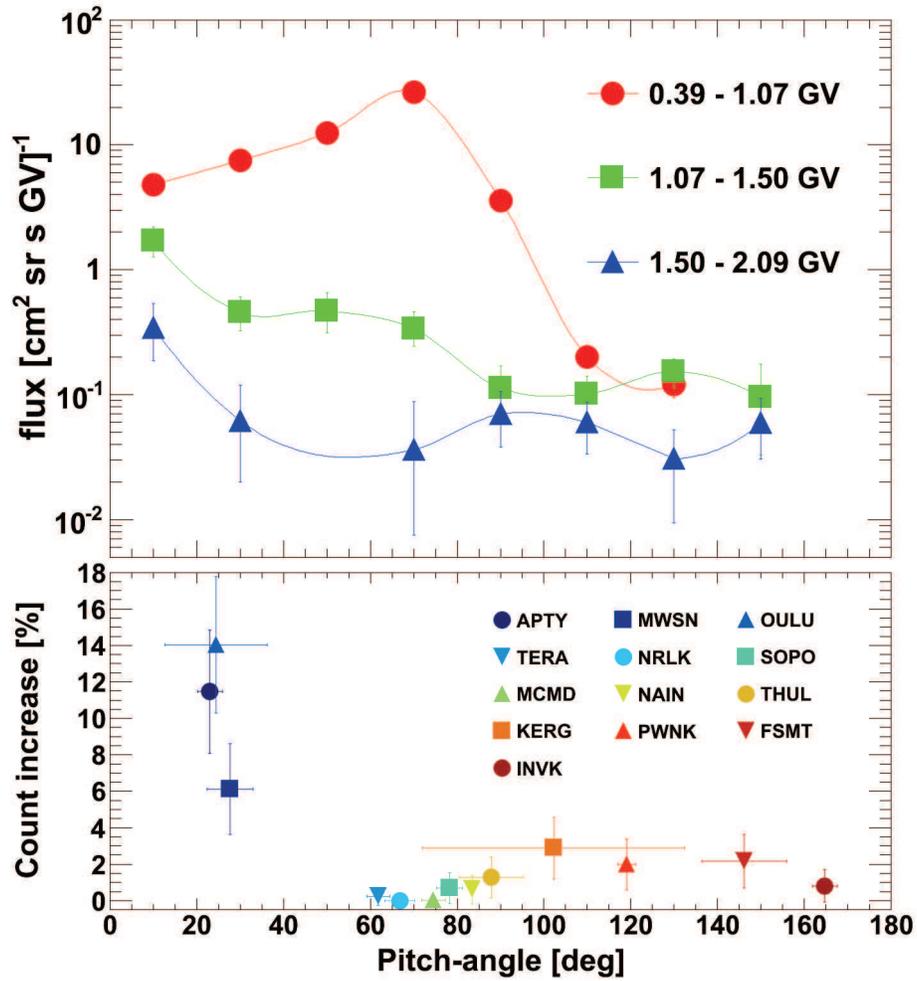}
\caption{(top) PAMELA's pitch angle distribution (background subtracted) in 3 rigidity ranges. The curves through the data are meant to guide the eye. Also shown is the world-wide neutron monitor pitch angle distribution averaged between 0158 to 0220 UT.  \label{fig3}}
\end{figure}

\clearpage

\begin{figure}
\epsscale{0.8}
\plotone{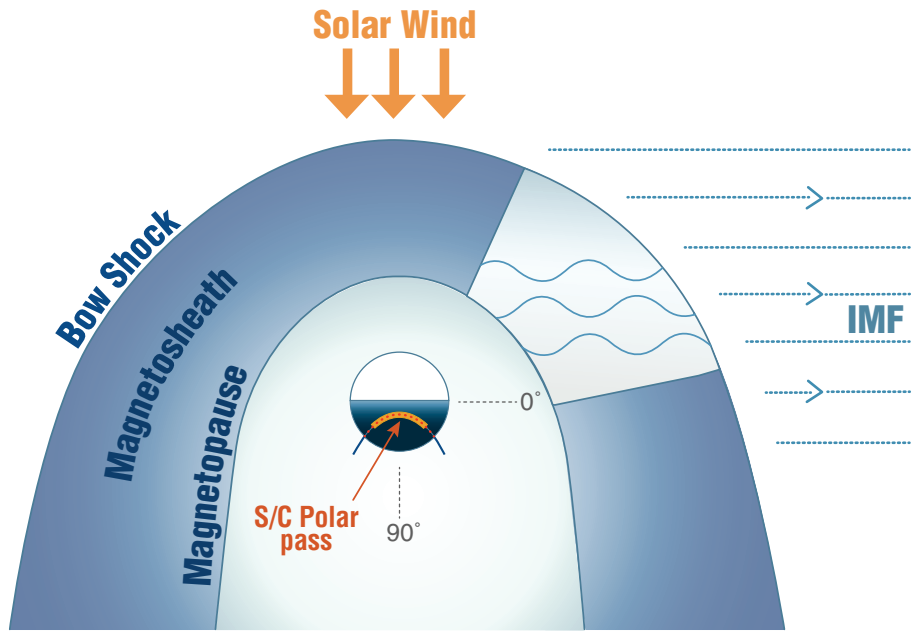}
\caption{Particle transport for the May 17$^{th}$ GLE through the Earth's magnetosphere. Due to the quasi-perpendicular nature of the IMF, mirror mode fluctuations are enhanced that redistribute the low-energy SEPs preferentially. The polar orbit of PAMELA is shown for context.    \label{fig4}}
\end{figure}

\clearpage

\end{document}